%%%%%%%%%%%%%%%%%%%%%%%%%%%%%%%%%%%%%%%%%%%%%%%%%%%%%%%%%%%%%%%%%%%%%%

% File : cone.tex

% Plain TeX source code for `Generalised Functions and Distributional
% Curvature of Cosmic Strings' (Clarke et al.) Submitted to GRQC by
% Jonathan P. Wilson (28 May 1996).

%%%%%%%%%%%%%%%%%%%%%%%%%%%%%%%%%%%%%%%%%%%%%%%%%%%%%%%%%%%%%%%%%%%%%

% Use A4 paper

\hsize=160mm
\vsize=235mm
\hoffset=0mm
\voffset=0mm

% This paper has been submitted to Classical and Quantum Gravity so
% originally used the IOP macros. This file is a free standing Plain TeX
% version which requires no additional macro files. The following macros
% are included for compatibility purposes.
%
\newcount\secno
\newcount\subno
\def\section#1{%
    \subno=0\global\advance\secno by 1 \bigbreak\leftline{\bf
    \number\secno. #1}\medbreak\noindent\ignorespaces}
\def\subsection#1{\global\advance\subno by 1
     \medbreak\leftline{\sl \number\secno.\number\subno. #1}\par
     \noindent\ignorespaces}
\def\definition#1{\bigbreak\noindent{\sl Definition #1}.\quad\ignorespaces}
\def\Remark#1{\bigbreak\noindent{\sl Remark #1}.\quad\ignorespaces}
\def\endproclaim{\par\bigbreak}
\let\Bbb=\bf
\let\bdi=\bf
\def\d{{\rm d}}
\def\e{{\rm e}}
\def\i{{\rm i}}

% define additional macros

\def\pfcase#1{\bigbreak\noindent{\sl Case #1}.\quad\ignorespaces}
%
% Use \displaystyle entries in \matrix
%
{\catcode`\@=11
\gdef\matrix#1{\null\,\vcenter{\normalbaselines\m@th
 \ialign{\hfil$\displaystyle{##}$\hfil&&\quad\hfil$\displaystyle{##}$\hfil\crcr
  \mathstrut\crcr\noalign{\kern-\baselineskip}
   #1\crcr\mathstrut\crcr\noalign{\kern-\baselineskip}}}\,}
}
\def\bmatrix#1{\left[\matrix{#1}\right]}

\def\crmsk{\cr\noalign{\medskip}}
%
% `black-board bold' symbols for reals, integers etc.
%
\def\Nat{{\Bbb N}} 
\def\Zint{{\Bbb Z}}
\def\Real{{\Bbb R}}
\def\Cplx{{\Bbb C}}
%
% miscellaneous symbols
%
\def\setabst#1#2{\left\{\, #1 \mid #2 \,\right\}} % set abstraction
\def\dirac#1{\delta^{(#1)}}      % Dirac delta function
\def\seq#1{\{#1\}}               % sequences
\def\const{{\rm constant}}       % constant
\def\bigO#1{O\left(#1\right)}    % O(r^n) notation
\def\norm#1{\left|#1\right|}     % norms
\def\Zsum#1{\sum_{#1\in\Zint}}   % sum over integers
\def\gCplx{{\bar\Cplx}}          % generalised numbers
\let\assoc=\vdash                % association
\let\weakequiv=\approx           % weak equivalence
\def\met#1#2{g_{#1#2}}           % metric
\def\hmet#1#2{h_{#1#2}}          % metric
\def\tempa#1{{#1\over x^2+y^2}}  % over r^2 
\def\supp{\mathop{\rm supp}\nolimits} % support of a function
%
% caligraphic letters
%
\def\calA{{\cal A}}
\def\calD{{\cal D}}
\def\calE{{\cal E}}
\def\calG{{\cal G}}
\def\calI{{\cal I}}
\def\calN{{\cal N}}
%
% bold letters
%
\def\ibold{{\bf i}}
\def\xbold{{\bdi x}}
\def\ybold{{\bdi y}}
\def\zbold{{\bdi z}}
\def\alphabold{{\bdi\alpha}}
%
% tilde symbols
%
\def\tildep#1{{\tilde #1}_\epsilon}
\def\wtildep#1{{\widetilde{#1}}_\epsilon}
\def\htildep{\tildep h}
\def\gtildep{{\tildep g}}
\def\gtilde{{\tilde g}}
\def\Rtilde{\tilde R}
\def\Rtildep{\tildep R}
\def\tildemet#1#2{\wtildep{\met{#1}{#2}}}
\def\tildehmet#1#2{\wtildep{\hmet{#1}{#2}}}

% BEGINNING OF THE PAPER ....

\font\bigbf=cmbx10 scaled \magstep3

{\baselineskip=1.7\baselineskip\bigbf
\centerline{Generalised Functions and Distributional Curvature}
\centerline{of Cosmic Strings}
}
\bigskip\medskip
\centerline{C J S Clarke, J A Vickers and J P Wilson}
\bigskip
\centerline{{\sl Department of Mathematics, University of Southampton,
Southampton, SO17~1BJ, UK}}
\medskip\bigskip

\noindent{\bf Abstract}\quad
A new method is presented for assigning distributional curvature, in
an invariant manner, to a space-time of low differentiability, using
the techniques of Colombeau's ``new generalised functions''.  The
method is applied to show that curvature of a cone is equivalent to a
delta function.  The same is true under small enough perturbations.

\vskip2\bigskipamount

% SECTION 1 : INTRODUCTION

\section{Introduction}
For some time it has been recognised that there is an important place
in relativity for metrics whose curvature has to be regarded as a
distribution; that is, the components of the Riemann tensor have to be
interpreted not as ordinary functions, but as functionals (Gel'fand
and Shilov, 1963) or as ideal limits of functions.  This has been
applied successfully to such examples as surface-distributions of
matter (Israel, 1966; Clarke and Dray, 1987) and to gravitational
radiation (Kahn and Penrose, 1971). In these cases, however, it is
possible to formulate the field equations so as to avoid operations
that are not well defined in ordinary distribution theory:
specifically, multiplying distributions, or multiplying, say, a
delta-function by a discontinuous function. The potential ambiguity of
such operations was established by Schwarz (1954), who pointed out
that it was impossible consistently to define an associative
multiplication on distributions, together with an operation of
differentiation, which coincided with the usual definition of these on
continuous and $C^1$ functions, respectively, and for which there was
a non-zero distribution $\delta$ satisfying $x\delta(x)=0$.

The curvature tensor is a non-linear function of the metric, so that
if one is to avoid these illegitimate operations, then there are
strong constraints on the sort of metric that can be considered.
Specifically, Geroch and Traschen (1987, Theorem 1) showed that in
order for the components of the Riemann tensor and its contractions to
be well defined as distributions, their singular parts have to have
support on a submanifold of dimension of at least three. This
unfortunately excludes many space-times which one expects to have a
distributional curvature, for example the cone-like space-time
$$\d s^2=-\d t^2+\d r^2+A^2r^2\d\phi^2+\d z^2, \qquad \norm{A}<1.\eqno(1)$$

Various attempts (Raju, 1982; Balasin and Nachbagauer, 1993) have been
made to give prescriptions for multiplying distributions in order to
overcome the problem posed by the nonlinearity of Einstein's
equations, but they have tended to depend on regularization procedures
whose invariance and general applicability were uncertain. There has
been until recently no general theory within which it could be shown
that the result was invariantly defined, independently of the
particular regularization adopted.

Our aim in this paper is to describe within the setting of relativity
theory the formalism, due to Colombeau (1983, 1990), in which
distributional curvatures can be rigorously and unambiguously defined.  We
will illustrate the method in the case of the conical metric~(1), where
one expects there to be a distributional energy momentum tensor of the
``thin string'' type (Kibble, 1976).  The illustration will be extended to
appropriately smooth perturbations of the metric. 

Roughly speaking, the method involves extending the Schwartz space of
distributions to a much larger space, whose elements we will call
generalised functions, within which the operations for computing the
curvature can always be defined. A subspace of the generalised
functions can then be defined whose elements correspond, in a many-one
manner, to classical distributions. For those metrics whose curvature
components lie in this subspace we can assign a well defined
distributional curvature. This procedure circumvents the result of
Schwarz quoted above by violating both the main conditions of
Schwartz' theorem: in the space of generalised functions
$x\delta(x)\neq 0$, and multiplication does not coincide with ordinary
multiplication for continuous functions (although it does for
$C^\infty$ functions).  These properties of generalised functions
(which would usually be regarded as undesirable) are resolved when the
generalised functions are mapped onto ordinary distributions, where
this is possible.  When this is done, $x\delta(x)$ corresponds to the
zero distribution;

% SECTION 2 : AN OVERVIEW OF COLOMBEAU'S GENERALISED FUNCTIONS

\section{An overview of Colombeau's generalised functions}
\subsection{Smoothing distributions}
As we shall be concerned both with space-time, locally $\Real^4$, and
with a 2-dimensional plane transverse to the conical singularity, it
will be convenient to work in $\Real^n$ for general $n$.  The basic
tool will be the operation of smoothing distributions.  Suppose $\Phi$
is a member of the space $\calD(\Real^n)$ of test functions: smooth
(i.e.\ $C^\infty$) $\Cplx$-valued functions on $\Real^n$ with compact
support; and that
$$ \int\Phi(\xbold)\, \d\xbold = 1. $$ 
Given $\epsilon>0$, we define
$$
\Phi^\epsilon(\xbold)={1\over\epsilon^n}
\Phi\left(\xbold\over\epsilon\right)
$$
so that $\Phi^\epsilon$ has a support scaled by $\epsilon$ and an
amplitude adjusted so that its integral is still unity.  If
$f:\Real^n\to\Cplx$ is a function, not necessarily continuous, then by
a smoothing of $f$ we mean one of the convolutions
$$
\tilde f(\xbold) := \int f(\ybold+\xbold)
 \Phi(\ybold) \, \d\ybold = \int f(\zbold)
 \Phi(\zbold-\xbold) \, \d\zbold.
$$
or
$$
\tilde f_\epsilon(\xbold) := \int f(\ybold+\xbold)
 \Phi^\epsilon(\ybold) \, \d\ybold = \int f(\zbold)
 \Phi^\epsilon(\zbold-\xbold) \, \d\zbold.
$$
(Smoothed functions will always depend implicitly on $\Phi$ as well as
on the explicit arguments $\xbold$ and $\epsilon$.)

Smoothing is defined in the same way for distributions, but with some
notational changes. A distribution $R$ is regarded as a $\Cplx$-valued
functional
$$ \calD(\Real^n)\ni\phi\mapsto (R,\phi)\in \Cplx $$
on the space $\calD(\Real^n)$ of test functions, and the convolution
is defined by
$$ \tilde R(\xbold) = (R,\Phi(\;.\;-\xbold)).$$
$$ \tilde R_\epsilon(\xbold) = (R,\Phi^\epsilon(\;.\;-\xbold)). $$

An intuitively plausible procedure for defining the product $Rf$ of, say,
a distribution $R$ and a discontinuous function $f$ would then be to
define the action of the product on a test function $\psi$ by first
defining the corresponding action of the product of the smoothed
quantities 
$\tildep R$ and $\tildep f$, and then taking the limit as the
smoothing is made progressively finer, with $\epsilon\to 0$:
$$ (Rf, \psi) = \lim_{\epsilon\to0}\int\tilde R_\epsilon(\xbold)\tildep
f(\xbold)\psi(\xbold) \, \d\xbold. $$
For example, if $R=\delta$, the Dirac $\delta$-function, and $f$ is
the Heaviside function, then this prescription yields the attractive
solution $\delta f=\delta/2$, {\it provided} that
$\Phi(-\xbold)=\Phi(\xbold)$. If one considers products involving more
complicated distributions, then the dependence of the answer on the
nature of $\Phi$ becomes more detailed. For example, writing $x^{-1}$
for the distribution defined by taking the Cauchy principal value in
integrals, 
$x^{-1}\delta=k\delta'$ where $k$ is a
$\Phi$-dependent constant.

A further complication arises in the context of relativity, where
one would like the results of multiplication to be invariant under 
$C^\infty$ coordinate changes. The underlying problem here is that the
operation of convolution depends on the linear structure of $\Real^4$
and so is not invariant under such changes. This suggests that
smoothing by convolution should be regarded as a special case of a
more general sort of smoothing, invariant under coordinate changes.

Colombeau's definition of generalised functions therefore starts from
a space $\calE_M(\Real^n)$ of functions depending on both the position
$x$ and a smoothing kernel $\Phi$, on which are imposed conditions
that reflect the special case of the smoothing of a distribution, but
which are more general.  Multiplication is defined on these pointwise.
This produces a space with some of the required properties, but in
which multiplication does not coincide with the ordinary
multiplication even for $C^\infty$ functions.  This is rectified by
defining an equivalence relation on $\calE_M(\Real^n)$ and passing to
a space $\calG$ of equivalence classes.  At the expense of minor
complications, the whole of the construction may be applied to a
domain which is a subset $\Omega$ of $\Real^n$, in which case we
denote the space of functions $\calE_M(\Omega)$ and the space of
equivalence classes $\calG(\Omega)$.

\subsection{Specification of the algebra}
We now give a precise definition of the space of generalised
functions $\calG(\Real^n)$.  We shall use standard multi-index notation
$$ \ibold=(i_1,\ldots,i_n), \qquad \norm\ibold=i_1+\cdots+i_n. $$
so
$$ \xbold^\ibold= {x_1}^{i_1} \ldots {x_n}^{i_n} $$
and
$$
 D^\ibold = {\partial^{\norm\ibold} \over \partial {x_1}^{i_1} \ldots
   \partial {x_n}^{i_n}}.
$$

An essential role in defining both the basic space of functions
$\calE_M(\Real^n)$ and the equivalence relation used to define the
final generalised functions is played by a classification of smoothing
kernels into subsets  $\calA_q(\Real^n)$.

\definition{1}
For $q\in\Nat$ we define $\calA_q(\Real^n)$ to be the set of functions
$\Phi\in\calD(\Real^n)$ such that
\item{(i)} $ \int_{\Real^n} \Phi(\xbold) \, \d\xbold=1$ \par
\item{(ii)} $ \int_{\Real^n} \Phi(\xbold) \xbold^\ibold \, \d\xbold=0
        \qquad \forall\ibold\in\Nat^n
        \quad \hbox{such that} \quad \norm\ibold \leq q $ \par
\endproclaim

We then define $\calE_M(\Real^n)$ in two steps. First, we set
$$
\calE(\Real^n)=\setabst{R:\calA_1\times\Real^n\to\Cplx}
{\hbox{$\xbold\mapsto R(\Phi,\xbold)$ is $C^\infty$}}
$$
This is an algebra (a vector space furnished with a multiplication)
under the operation of pointwise multiplication
$$
(RS)(\Phi,\xbold) := R(\Phi,\xbold) S(\Phi,\xbold).
$$
The derivatives of these functions with respect to $\xbold$ will be
denoted by $D^{\alphabold} R(\Phi,\xbold)$.

The second step is as follows:

\definition{2}
The subalgebra $\calE_M(\Real^n)$ (functions of moderate growth), of
$\calE(\Real^n)$ is defined to be the set of functions $R$ such that
for all compact $K\subseteq\Real^n$ and for all $\alphabold\in\Nat^n$,
there is some $N\in\Nat$ such that: If $\Phi\in\calA_{N}$, $\exists
c,\eta>0$ such that
$$
\norm{ D^{\alphabold} R(\Phi^\epsilon,\xbold) } \leq c\epsilon^{-N} 
\qquad (x\in K,\ 0<\epsilon<\eta)
$$
\endproclaim

It is easily verified that, for any distribution $T$ with support in
$\Real^n$, the corresponding smoothed function $\tilde T$ is in
$\calE_M(\Real^n)$.

Finally, we need to define the equivalence relation that will give a
multiplication that coincides with the usual one on $C^\infty$
functions. The key idea is to note that there are two different ways
in which ordinary functions $f$ can be mapped into elements of
$\calE_M(\Real^n)$: we can (for any continuous $f$) form by smoothing
the element
$$
R_f(\xbold,\Phi)= \int f(\ybold+\xbold)
 \Phi(\ybold) \, \d\ybold
$$
or (for $C^\infty$ functions $f$ only) we can form the element
$$
S_f(\xbold, \Phi) = f(\xbold).
$$
Multiplication of the $S_f$ trivially coincides with ordinary
multiplication of the functions $f$, whereas this does not hold for
the $R_f$. Colombeau therefore defines an equivalence relation that
identifies $S_f$ with $R_f$. This ensures that for $C^\infty$
functions there is a single mapping of functions into generalised
functions (equivalence classes), which coincides with the mapping by
smoothing used for distributions, and on which multiplication
coincides with ordinary multiplication for $C^\infty$ functions. The
equivalence is defined by means of an ideal $\calN(\Real^n)$ (i.e.\ a
subalgebra such that if $R\in \calN(\Real^n)$ and $S\in
\calE_M(\Real^n)$ then $RS\in\calN(\Real^n)$) so that elements
differing by a member of $\calN(\Real^n)$ are identified.

\definition{3}
The ideal $\calN(\Real^n)$, of $\calE_M(\Real^n)$ is defined to be the set
of functions $R$ such that for all compact $K\subseteq\Real^n$ and for all
$\alphabold\in\Nat^n$, there is some $N\in\Nat$ and some increasing and
unbounded sequence $\seq{\gamma_n}$ such that: If $\Phi\in\calA_{q}$, for
$q\geq N$, $\exists c,\eta>0$ such that
$$
\norm{ D^\alphabold R(\Phi^\epsilon,\xbold) } \leq c\epsilon^{\gamma_q-N} 
\qquad (x\in K,\ 0<\epsilon<\eta)
$$
\endproclaim

For the smoothed functions used here, $\gamma_q$ will be simply $q$;
the more general form is used so as to ensure invariance under
coordinate changes.

The key result is then that, for $C^\infty$ functions $f$, $R_f-S_f$
is in $\calN(\Real^n)$.

\definition{4 (Generalised functions)}
The algebra of generalised functions is defined to be
$$
\calG(\Real^n)={\calE_M(\Real^n) \over \calN(\Real^n)}.
$$
\endproclaim

The generalised function corresponding to the product of two
distributions $T$ and $U$, or of a distribution $T$ and a continuous
function $f$, is then taken to be the equivalence class of the product
$\tilde T\tilde U$, or $\tilde T\tilde f$ respectively.

\subsection{Partial correspondence with distributions}
In order to associate a generalised function with a distribution we need
to give a meaning to the expression $\int G(\xbold) \phi(\xbold)\,
\d\xbold$ with $G$ a generalised function and $\phi$ a test function. 
Colombeau therefore introduces a concept of the integral of a generalised
function. 

Given a generalised function $G=[R(\Phi,\xbold)]$, given as the
equivalence class of an element $R\in\calE_M(\Real^n)$, the quantity
$$
\rho := \int_{\Real^n} R(\Phi,\xbold) \, \d\xbold
$$
is a function of $\Phi$ that depends on the choice of $R$. More
specifically, it is a function $\rho:\calA_1\to\Cplx$ satisfying
the condition:

{\parindent=30pt\narrower
{there is some $N\in\Nat$ such that: If
$\Phi\in\calA_N$, $\exists c,\eta>0$ such that}
$$
\norm{ \rho(\Phi^\epsilon) } \leq c\epsilon^{-N} 
\qquad (0<\epsilon<\eta)
$$
\par}

\definition{5}
The algebra $\calE_M$ is defined to be the set of
functions $\rho$ satisfying the above condition.
\endproclaim

Different choices of the representative $R$ for $G$ result in functions
$\rho$ differing by a function in a subalgebra $\calI$ playing the same
role as $\calN(\Real^n)$:

\definition{6}
The ideal $\calI$, of $\calE_M$ is defined to be the set of functions
$R$ such that there is some $N\in\Nat$ and some increasing and
unbounded sequence $\seq{\gamma_n}$ such that: If $\Phi\in\calA_q$ for
$q\geq N$, $\exists c,\eta>0$ such that
$$
\norm{ \rho(\Phi^\epsilon) } \leq c\epsilon^{\gamma_q-N} 
\qquad (0<\epsilon<\eta)
$$
\endproclaim

Thus if we take the quotient of $\calE_M$ by $\calI$ we will obtain an
equivalence class that is independent of the representative $R$, and
that can therefore be regarded as a value of the integral of the
generalised function $G$. Whereas the integral of an ordinary complex
function is a complex number, an element of this quotient is called
a generalised complex number. 

\definition{7 (Generalised Numbers)}
The algebra of generalised numbers is defined to be
$$
\gCplx={\calE_M \over \calI}.
$$
\endproclaim

For each ordinary ordinary complex number $z$ we can form the
generalised complex number $\bar z$ defined as the equivalence
class $[\rho_z]$ of the constant function $\rho_z(\Phi)=z$. It follows
then that $z$ is {\sl associated to} $\bar z$, in the following sense:

\definition{8}
We say that $\bar z\in\gCplx$ is associated to the classical
number $z\in\Cplx$ (written as $\bar z \assoc z$) if there is some
representative $\rho\in\calE_M$ of $\bar z$ such that for $\Phi\in\calA_q$
with $q$ large enough, $$\lim_{\epsilon\to0}\rho(\Phi^\epsilon)=z.$$
\endproclaim

\definition{9}
We say that $\bar z_1, \bar z_2 \in\gCplx$ are associated to
each other if and only if $\bar z_1-\bar z_2 \assoc 0\in\Cplx$
\endproclaim
Note that $\bar z\assoc z$ does not necessarily imply $\bar z =
[\rho_z]$. It is, however, this weaker notion of association that
turns out to be the important one in assigning corresponding
distributions to generalised functions. To do this, we first make the
following:

\definition{10 (Weak equivalence)}
We say that $G_1, G_2 \in \calG(\Real^n)$ are associated to each other
(written as $G_1 \weakequiv G_2$) if and only for each $\Psi\in\calD$,
$$
\int_{\Real^n} (G_1(\xbold)-G_2(\xbold)) \Psi(\xbold) \, \d\xbold
\assoc 0 \in\Cplx.
$$
\endproclaim

Then we say that a generalised function $G$ corresponds to a
distribution $T$ if $G\weakequiv \tilde T$. For physical purposes, we
are interested in those generalised functions, defined by replacing the
terms in the definition of the Riemann tensor components by smoothed
functions and distributions, which correspond to distributions.

% SECTION 3 : SMOOTHING THE CONE

\section{Smoothing the cone}
In this section we will look at the curvature of the spacetime given by
the metric~(1). In this example the singular part of
the curvature arises because of a conical singularity in the
2-dimensional plane transverse to the axis. We will therefore examine
the curvature of the 2-dimensional metric
$$
\d s^2=\d r^2+A^2 r^2 \d\phi^2. \qquad \norm{A}<1.
\eqno(2)
$$
Since this is the metric of a surface the above example has the advantage
that the curvature may be given by a scalar quantity, the Gaussian
curvature, rather than having to use the curvature tensor. This
simplifies the notation and the details of the smoothing but plays no
essential role in the calculation. We explore the implications of this
further in the final section. 

We want to think of the cone metric as being defined on the whole of
$\Real^2$ so that we start by expressing~(2) in coordinates which are
regular at the origin.  In Cartesian coordinates ($x=r\cos\phi$,
$y=r\sin\phi$) the metric is given by
$$
\met{a}{b}={1\over2}(1+A^2)\delta_{ab} +{1\over2}(1-A^2)h_{ab}
%\eqno(*)
$$
where
$$
  h_{ab} = \bmatrix{
             \tempa{x^2-y^2} & \tempa{2xy}      \cr
             \tempa{2xy}     & \tempa{y^2-x^2}  \cr }
$$
Note that $h_{ab}$ represents the `singular' i.e.\ the non-smooth part
of the metric.

% introduce smoothed metric

We now wish to regard the components of the metric as elements of
$\calE_M(\Real^2)$ so that for  $\Phi\in\calA_1(\Real^2)$, 
$$
 \tildemet{a}{b}(x,y) = \int_{\Real^2} \met{a}{b}(u,v)
   \Phi^\epsilon(u-x,v-y) \,\d u\,\d v,
$$
are smooth functions which we regard as the components of a smoothed
metric. We can therefore calculate the Ricci curvature $\Rtildep$ of
this metric which may also be regarded as an element of
$\calE_M(\Real^2)$.  In this section we show that the curvature is
given by a delta function in the sense that
$[\Rtilde\sqrt\gtilde]\weakequiv 4\pi(1-A)\dirac2$.

\Remark We shall regard the delta function as a scalar density, and the
functions in $\calD$ as scalars.  Thus a factor of $\sqrt\gtilde$ is
inserted to make the left hand side a density. \endproclaim

It is important to note that although the calculation is carried 
out using the Cartesian components
of the metric for convenience, the definitions of the Colombeau
algebra ensure that the result does not depend upon the
choice of coordinates used provided they are related by a smooth
transformation. 

In order to calculate $\tildemet{a}{b}$ we first note that constant
functions are unchanged on smoothing so that
$$
\tildemet{a}{b}(x,y) = {1\over2}(1+A^2) \delta_{ab} 
  + {1\over2}(1-A^2) \tildehmet{a}{b}
   \eqno(3)
$$
We next note that 
$$
\tempa{x^2-y^2}+\i\tempa{2xy}=\e^{2\i\phi}
$$
So that we may calculate $\tildehmet{a}{b}$ by smoothing the complex valued
function $h(x,y)=\e^{2\i\phi}$ to obtain
%
%
% Here is the smoothing of h(x,y). This bit will be reused in later
% calculations. e.g. 4.D cones/cosmic strings.
%
%
$$
\htildep(x,y) = {1\over\epsilon^2} \int_{\Real^2} h(u+x,v+y)
 \Phi (u/\epsilon,v/\epsilon) \, \d u\, \d v
$$
In order to calculate the integral we express it in terms of polar
coordinates $(r,\phi)$ and expand $\Phi$
in terms of circular harmonics. Writing
$$
\Phi(r,\phi) = \Zsum{n} \Phi_n(r) \e^{\i n\phi}
$$
we have
$$
\htildep(r,\phi) = {1\over\epsilon^2} \Zsum{n}
 \int^\infty_0 I_n( r',r,\phi) \Phi_n( r'/\epsilon)  r'\,\d r'
$$
where
$$
I_n=\int^{2\pi}_0 { (r\e^{\i\phi}+ r' \e^{\i\phi'})^2 \over r^2
 +  r'^2 +2r r'\cos(\phi-\phi') } \e^{\i n\phi'} \, \d\phi'.
$$
Setting $w=\e^{\i\phi}$ and $z=\e^{\i\phi'}$, we may integrate out
$\phi'$ by considering the integral of the complex function
$$
\Gamma(z)=-\i \left( r'z + rw \over rz +  r'w \right) wz^n
$$
around the circular contour $\kappa:\norm{z}=1$.

If $ r'<r$, a single pole occurs at $z=- r'w/r$ which has a residue of
$-\i(- r'/r)^n(1- r'^2/r)\e^{\i(n+2)\phi}$.

If $n=-1$, a single pole will occur at $z=0$ which has a residue of
$-\i(r'/r)\e^{\i\phi}$.

If $n\leq-2$, a $\norm{n}$-fold pole will occur at $z=0$ which has a residue of
$\i(- r'/r)^n(1- r'^2/r)\e^{\i(n+2)\phi}$.

Using the residue theorem, we find that
$$
I_n(r',r,\phi)= \cases{
       2\pi \left( 1-{ r'^2\over r^2} \right)
       \left(-{ r'\over r} \right)^n \e^{\i(n+2)\phi}
    &  if $ r'<r$ and $n\geq0$, \crmsk
      -2\pi \left( 1-{ r'^2\over r^2} \right)
       \left(-{ r'\over r} \right)^n \e^{\i(n+2)\phi} 
    &  if $ r'>r$ and $n\leq-2$, \crmsk
       2\pi \min\{r'/r,r/r'\} \e^{\i\phi}, 
    &  if $n=-1$, \crmsk
  0 &  otherwise. \cr 
}
$$
So if we expand $\htildep(r,\phi)$ as
$$
\htildep(r,\phi) = \Zsum{n} H_{n,\epsilon}(r) \e^{\i n\phi},
$$
we obtain
$$
H_{n,\epsilon}(r)=\cases{
     2\pi(-1)^n \int^{r/\epsilon}_0
     \left( \epsilon  r' \over r \right)^{n-2}
     \left( 1-{\epsilon^2  r'^2 \over r^2} \right)
     \Phi_{n-2}( r')  r'\, \d r',
   & if $n\geq2$, \crmsk
     2\pi \int^{r/\epsilon}_0 \left( \epsilon  r' \over r \right)
     \Phi_{-1}( r')  r'\,\d r' +2\pi\int^\infty_{r/\epsilon}
     \left( r \over \epsilon  r' \right) \Phi_{-1}( r')  r' \,\d r',
   & if $n=1$, \crmsk
    -2\pi(-1)^n \int^\infty_{r/\epsilon}
    \left( \epsilon  r' \over r \right )^{n-2}
    \left( 1-{\epsilon^2  r'^2 \over r^2} \right)
    \Phi_{n-2}( r')  r'\, \d r',
  & if $n\leq0$, \cr
}
$$
We next obtain estimates for the $r$ dependence of
$H_{n,\epsilon}(r)$. We will be interested in the behaviour both for
small and large $r/\epsilon$.  Let $
R_0=\sup\setabst{r}{\hbox{$\norm{\Phi_n(r)}>0$ for some $n\in\Zint$}}$
then we consider exterior case $r>\epsilon R_0$ first.

% This is the exterior bit ....
%
%
\pfcase{1 ($r>\epsilon R_0$)}.

In polar coordinates the condition that $\Phi\in\calA_q$ may be expressed as
$$
\eqalignno{
  &  \int^\infty_0 \Phi_0(r) r \, \d r =1 
  &  (4)\cr
  &  \int^\infty_0 r^{c+1} \Phi_n(r) \,\d r=0, \qquad
     {\norm{n}\leq c\leq q \atop \hbox{$c+n$ is even}}
  &  (5)\cr
  }
$$

Fixing $r>0$ and choosing $\epsilon<r/ R_0$, we find that by~(4)
$$
 H_{2,\epsilon}(r)= 1-2\pi \int^\infty_{r/\epsilon}  r'\Phi_0( r') \, \d r'
      -2\pi {\epsilon^2 \over r^2}
       \int^{r/\epsilon}_0  r'^3 \Phi_0( r') \, \d r' 
$$
Note that the first integral gives zero because
$r/\epsilon> R_0$.  So that
$$
 H_{2,\epsilon}(r)= 1+ O(\epsilon^2/r^2)
$$
However if $\Phi\in\calA_q$ with $q \geq 2$, we may use~(5)
to express $H_{2,\epsilon}$ as
$$
  H_{2,\epsilon}(r)=1-2\pi \int^\infty_{r/\epsilon}  r' \Phi_0( r')
  \,\d r' + 2\pi{\epsilon^2\over r^2} \int^\infty_{r/\epsilon}
   r'^3 \Phi_0( r') \,\d r'
$$
in which both integrals vanish for $\epsilon<r/ R_0$.  A similar 
calculation shows that for $ n=1$ and for $n > 2$ 
$$
H_{n,\epsilon}(r) = O(\epsilon^n/r^n)
$$
with a coefficient that vanishes if $\Phi \in \calA_q$ with $q \geq n$.

On the other hand, by the definition of $R_0$ 
$$
H_{n,\epsilon}(r) = 0 \quad \hbox{for} \quad n<0\;.
$$

Putting these results together we have in this case for $\Phi\in\calA_q$,
$$
\htildep(r,\phi) = \e^{2\i\phi}+\bigO{\epsilon^{q+1} \over r^{q+1}} \;.
$$

\pfcase{2 ($r<\epsilon  R_0$)}
%
%
% And the interior bit.
%
%
In the following $C_1$, $C_2$ etc.\ denote positive constants.

For $n\geq2$ or $n\leq-3$ we use the fact that $\norm{\Phi} \leq K=
\sup\norm{\Phi}$ to obtain
$$
   \norm{H_{n,\epsilon}} \leq 2\pi K {r^2\over\epsilon^2}
   \norm{{1\over n}-{1\over n+2}} =O(r^2/\epsilon^2)
$$
The cases $n=-2,-1,0,1$ are more delicate. Here we expand
$H_{n,\epsilon}$ in a Taylor series (with remainder). For the case of
$H_{0,\epsilon}$ we have
$$
\eqalign{
H_{0,\epsilon}(r)   &= 2\pi \int^\infty_{r/\epsilon} \left(
            1-{r^2\over\epsilon^2 r'^2} \right)
            \Phi_{-2}( r') \, \d r' \cr
H_{0,\epsilon}'(r)  &= {2\pi\over\epsilon} \int^\infty_{r/\epsilon}
            \left( -2r\over\epsilon r'\right)
            \Phi_{-2}( r') \, \d r' \cr
H_{0,\epsilon}''(r) &= {-2\pi\over\epsilon^2} \int^\infty_{r/\epsilon}
            {2\over r'} \Phi_{-2}( r') \, \d r' 
          + {2\pi\over\epsilon} \bigl[ 2\Phi_{-2}(r/\epsilon) \bigr] \cr
}
$$
and so
$$
\eqalign{
   H_{0,\epsilon}(0)   &= C_1\neq0 \cr
   H_{0,\epsilon}'(0)  &= 0 \cr
   H_{0,\epsilon}''(0) &= {C_2\over\epsilon^2} \cr
}
$$
giving $H_{0,\epsilon}(r)=C_3+O(r^2/\epsilon^2)$. A similar method gives
$$
\eqalign{
   H_{1,\epsilon}(r)    &= O(r/\epsilon), \cr
   H_{-1,\epsilon}(r) &= O(r/\epsilon), \cr
   H_{-2,\epsilon}(r) &= O(r^2\epsilon^2). \cr
}
$$
Therefore in this case we have
$$
  \htildep(r,\phi) = \alpha_0+\alpha_1 {r\over\epsilon} \e^{-\i\phi}
   + \alpha_2 {r\over\epsilon} \e^{\i\phi} +O(r^2/\epsilon^2)
$$
where $\alpha_n$ are constants.

% Display the smoothed metric.

The smoothed metric may now be expressed by~(3), as
$$
\def\tempa#1{{#1\over x^2+y^2}}
  \tildehmet{a}{b}=\bmatrix{
                  \tempa{x^2-y^2} & \tempa{2xy}     \cr
                  \tempa{2xy}     & \tempa{y^2-x^2} \cr
             } + \bigO{ \epsilon^{q+1} \over (x^2+y^2)^{q+1\over2}},
$$
for $r>\epsilon R_0$, and
$$
  \tildehmet{a}{b} = \bmatrix{
                  \beta_1+ {1\over\epsilon} (\beta_3x+\beta_4y) & 
                  \beta_2+ {1\over\epsilon} (\beta_5x+\beta_6y) \cr
                  \beta_2+ {1\over\epsilon} (\beta_5x+\beta_6y) &
                 -\beta_1- {1\over\epsilon} (\beta_3x+\beta_4y) \cr
                    } + \bigO{ x^2+y^2 \over \epsilon^2}
$$
for $r<\epsilon R_0$.

% Calculate Ricci tensor.

We may now calculate the the Ricci scalar $\Rtildep$ of the smoothed
metric.
$$
\Rtildep=\cases{
    O(1/\epsilon^2)  & 
    if $r<\epsilon  R_0$ \cr
    O\left( \epsilon^{q+1} \over r^{q+3} \right) &
    if $r>\epsilon  R_0$ \cr
}
$$
Note that if in Definition 3 we only consider compact sets $K$ not
containing the origin, then the second case in the above equation
holds for small enough $\epsilon$, and $\Rtildep$ then satisfies the
conditions for membership of $\calN$. This can be interpreted as
meaning that the curvature is concentrated at the origin.

We now show that
$[\Rtilde\sqrt\gtilde] \weakequiv 4\pi(1-A) \dirac2$.  It is sufficient to
show that for each $\Psi\in\calD(\Real^2)$,
$$
  \lim_{\epsilon\to0} \int_K \Rtildep\sqrt\gtildep(x,y)
  \Psi(x,y) \,\d x\,\d y = 4\pi(1-A)\Psi(0,0)
$$
for all $\Phi\in\calA_m$ for some $m\in\Nat^+$, where $K=\supp{\Psi}$.

By the mean value theorem we may express the left hand side as
$\Psi(0,0) I_1 + I_2$ where
$$
\eqalign{
   I_1 &= \int_K \Rtildep\sqrt\gtildep \,\d x\,\d y \cr
   I_2 &= \int_K \Rtildep\sqrt\gtildep r
          {{\partial\Psi} \over {\partial r}}(\xi x,\xi y) \,\d x\,\d y \cr
}
$$
for some $\xi\in[0,1]$. Letting
$$
\eqalign{
C &= \sup \left\{ {{\partial\Psi} \over {\partial r}}
  (\xi x,\xi y) \right\}  \cr
B_\epsilon &= \setabst{ (x,y)\in \Real^2}
  { (x^2+y^2)^{1/2}<\epsilon  R_0} \cr
}
$$
we may write
$$
 \norm{I_2} \leq C\norm{\int_{B_\epsilon} \Rtildep\sqrt\gtildep
                  r\,dx\,dy} + C\norm{\int_{K-B_\epsilon} 
                  \Rtildep\sqrt\gtildep r\,\d x\,\d y}       
$$
For the first integral
$$
\norm{\int_{B_\epsilon} \Rtildep\sqrt\gtildep r\,\d x\,\d y} \leq
C_4 {{(\epsilon R_0)^3} \over {\epsilon^2}}
$$
and for the second integral
$$
\norm{\int_{K-B_\epsilon} \Rtildep\sqrt\gtildep r\,\d x\,\d y} \leq
C_5\epsilon^{q+1}\left({1 \over {(\epsilon R_0)^q}}-{1 \over
{{R_K}^q}}\right)
$$
where $R_K$ is the maximum radius of $K$.

This gives
$$
\norm{I_2} \leq  C_6\epsilon
$$

To calculate $I_1$ we let 
$D=\setabst{(x,y)}{x^2+y^2\leq \mu^2}$ be a disc such that $D\subseteq K$ 
and split up the integral into two parts.
$$
I_1 = \int_{K-D} \Rtildep\sqrt\gtildep \,\d x\,\d y
      +\int_{D} \Rtildep\sqrt\gtildep \,\d x\,\d y
$$
For the first integral we have
$$
\int_{K-D} \Rtildep\sqrt\gtildep \,\d x\,\d y = \bigO{\epsilon^{q+1}/\mu^{q+1}}
$$
So that in the limit as $\epsilon \to 0$ the only non-zero
contribution comes from the integral over the disc. To calculate this
we apply the Gauss-Bonnet theorem to convert it into an integral
around the boundary.
$$
{1\over2}\int_D \Rtildep\sqrt\gtildep \,\d x\,\d y
  = 2\pi - \int_{\partial D} \kappa_\gtildep \,\d s
\eqno(6)
$$
where $\kappa_\gtildep$ is the geodesic curvature of the smoothed metric.
Since $D$ lies in the exterior region for small enough $\epsilon$ we get
$$
\int_{\partial D} \kappa_\gtildep \,\d s= 2\pi A+\bigO{{{\epsilon^{q+1}}
\over {\mu^{q+1}}}}
$$
so that
$$
I_1=4\pi -4\pi A +\bigO{{{\epsilon^{q+1}} \over {\mu^{q+1}}}}
$$
Therefore
$$
\lim_{\epsilon\to0} (\Psi(0,0) I_1+I_2) = 4\pi(1-A)
$$ 
as claimed.

% SECTION 4 : PERTURBATIONS.

% to be revamped sometime.

\section{Perturbations}
If we introduce a perturbation on the metric that is small enough, we
still obtain the same conclusion. Let us suppose we introduce an
$r^2$ perturbation so that the metric takes the following form.
$$
\d s^2= \d r^2 + A^2 r^2(1+kr^2) \d \phi^2
\qquad A,k=\const
$$
In Cartesian coordinates, the metric may be written as
$$
\met{a}{b}={1\over2}\bigl(1+A^2(1+kr^2)\bigr)\delta_{ab}
          +{1\over2}(1-A^2)h_{ab}
          -{1\over2}k A^2l_{ab}
$$
with the additional perturbation term being
$$
l_{ab} = \bmatrix{ x^2-y^2 & 2xy \cr 2xy & y^2-x^2 \cr }.
$$
Since a smooth function may be regarded as a generalised function with
no $\Phi$ dependence, an element of $\calE(\Real^2)$ equivalent to the
smoothed metric may be defined by
$$
 \tildemet{a}{b}(x,y) = {1\over2}(1+A^2+A^2 k r^2) \delta_{ab}
 + {1\over2}(1-A^2) \widetilde{h_{ab}}-{1\over2}A^2k {l_{ab}}
$$

We now consider the Ricci scalar of this metric. The main difference
between this case and a flat cone is that the curvature does not
vanish over away from the origin. In fact
$$
R=-2k {3+2kr^2 \over (1+kr^2)^2 }.
$$

Using the smoothed metric,
$$
\Rtildep=\cases{
    O(1/\epsilon^2)  & 
    if $r<\epsilon  R_0$ \cr
    R + O\left( \epsilon^{q+1} \over r^{q+3} \right) &
    if $r>\epsilon  R_0$ \cr
}
$$

We might expect to find that, in the sense of generalised
functions, $[\Rtilde\sqrt\gtilde] \weakequiv R\sqrt{g}+4\pi(1-A)
\dirac2$. We can verify this by performing a calculation similar to the
one above for the flat cone, but by replacing $\Rtilde\sqrt\gtilde$ by
$\Rtilde\sqrt\gtilde -R\sqrt{g}$ throughout.

The calculation of $I_2$ is unchanged. To calculate $I_1$ we write the
integral as
$$
\eqalign{
I_1 &= \int_D \Rtildep\sqrt\gtildep \,\d x\,\d y
       - \int_D R\sqrt{g} \,\d x\,\d y
       +\int_{K-D} (\Rtildep\sqrt\gtildep-R\sqrt{g}) \,\d x\,\d y \cr
    &= 4\pi -4\pi A {1+2k\mu^2 \over (1+k\mu^2)^{1/2}} + 4\pi A \left[
      {1+2k\mu^2 \over (1+k\mu^2)^{1/2}}-1 \right] + \bigO{ \epsilon^{q+1}
       \over \mu^{q+1} }. \cr
}
$$
where we have applied the Gauss-Bonnet theorem~(6) to evaluate the first
integral, and explicitly computed the second.  Therefore
$$
\lim_{\epsilon\to0} (\Psi(0,0) I_1+I_2) = 4\pi(1-A).
$$ 
and hence
$$
[\Rtilde\sqrt\gtilde] \weakequiv R\sqrt{g}+4\pi(1-A)\dirac2
$$
It is clear from the above that replacing $kr^2$ by a more general
perturbation of the form $k(r)r^2$ where $k(r)$ is a smooth function
would give the same result. Indeed any perturbation for which $l_{ab}$
is a smooth function would give a similar result.

% SECTION 5 : CONCLUSION

\section{Conclusion}
We have shown that the curvature of a 2-dimensional cone at $r=0$,
when regarded as a generalised function is weakly equivalent to the
delta distribution and the conclusion remains the same for a suitably
perturbed metric. To simplify the presentation we have confined our
attention to the two dimensional case however similar results hold for
the case of the the 4-dimensional cone-like space-time~(1).  In the
four dimensional case one must use a more general smoothing function
with both $t$ and $z$ dependence to calculate the transverse
components of $\tildemet{a}{b}$. However after integrating out in the
$t$ and $z$ directions the calculation is much the same. One may then
calculate the curvature tensor ${\tilde R}^a_{\phantom{a}bcd}$ of the
smoothed metric and regard the components as elements of
$\calE_M(\Real^4)$.  The only terms that might correspond to
distributions are those which are the sectional curvatures of the
2-surface transverse to the string.  By applying the methods of \S3
but using a holonomy argument rather than the Gauss-Bonnet theorem to
calculate the final integral one may show that these terms regarded as
generalised functions are weakly equivalent to delta functions, thus
confirming the results of Vickers (1987).

Currently work is in progress to extend these calculations to more
general four dimensional space-times in general relativity which admit
quasi-regular singularities. One result would be to derive a new
definition of the mass of a string per unit length, which Geroch and
Traschen showed was ambiguous when defined by limiting sequences of
metrics.

\begingroup

\bigbreak\vbox{%
\leftline{\bf References}\medbreak
\frenchspacing
\parskip=\smallskipamount
\parindent=0pt

\def\article#1#2#3#4#5#6{#1 #6 #2 {\it#3} {\bf#4} #5 \par}
\def\book#1#2#3#4{#1 #4 {\it#2} #3 \par}

\article{Balasin~H and Nachbagauer~H}
        {What curves the Schwarzschild Geometry}
        {Class. Quantum Grav.}
        {10}{2271--2278}{1993}

\book{Biagioni~H~A}
     {A nonlinear theory of generalised functions}
     {Springer-Verlag}
     {1990}

\article{Clarke~C~J~S and Dray~T}
        {Junction conditions for null hypersurfaces}
        {Class. Quantum Grav.}
        {4}{265--275}{1987}

\article{Colombeau~J~F}
        {A multiplication of distributions}
        {Journal of Mathematical analysis and applications}
        {94}{96--115}{1983}

\article{Colombeau~J~F}
        {A multiplication of distributions}
        {Bulletin of the American Mathematical Society}
        {23}{251--268}{1990}

\article{Geroch~R~P and Traschen~J}
        {Strings and other distributional sources in general relativity}
        {Phys. Rev~D}
        {38}{1017--1031}{1987}

\article{Israel~W}
        {Singular hypersurfaces and thin shells in General Relativity}
        {Nuovo Cimento B}
        {44}{1}{1966}

\article{Khan~K~A  and Penrose~R}
        {Scattering of two impulsive gravitational plane waves}
        {Nature}
        {229}{185--186}{1971}

\article{Kibble~T~W~B}
        {Topology of cosmic domains and strings}
        {J.\ Phys.\ A.}
        {9}{1387--1398}{1976}

\article{Raju~C~K}
        {Junction conditions in general relativity}
        {J.\ Phys.\ A}
        {15}{1785--1797}{1982}

\article{Schwarz~L}
        {Sur l'impossibilit\'e de la multiplication des distributions.}
        {C. R. Acad. Sci. Paris}{239}{847--848}{1954}

\article{Vickers~J~A}
        {Generalised Cosmic Strings}
        {Class. Quantum Grav.}{4}{1--9}{1987}

}\endgroup

\bye